\documentclass[conference]{IEEEtran}

\IEEEoverridecommandlockouts                             
\usepackage[mode=tex]{standalone} 

\usepackage{cite}
\usepackage{color}

\usepackage{verbatim}
\usepackage{booktabs}
\usepackage{amsmath}
\usepackage{balance}

\usepackage{graphicx}
\usepackage{tabu}
\usepackage[font={footnotesize}]{caption}
\usepackage{array, ltablex, multirow,ragged2e}
\usepackage{hyperref}
\usepackage{longtable}
\usepackage{mdframed}

\usepackage{subfigure}
\usepackage{dsfont}
\usepackage{float}
\usepackage{algorithm}
\usepackage[noend]{algpseudocode}
\usepackage{tikz,pgfplots,tkz-graph}
\usepgfplotslibrary{fillbetween}
\usepgfplotslibrary{groupplots}
\makeatletter
\def\BState{\State\hskip-\ALG@thistlm}
\makeatother

\usepackage{amsmath} 
\usepackage{amssymb}  
\usepackage{amsthm}
\usepackage{bm}
\theoremstyle{plain}

\theoremstyle{remark}

\theoremstyle{plain}

\graphicspath{{../figures/}}

\pgfplotsset{compat=newest}%
\usetikzlibrary{positioning,matrix,shapes.multipart,shapes.misc,spy}%
\tikzstyle{annotation}=[fill=white]%

\newcommand{\figwidth}{8.8cm}
\newcommand{\figheight}{7.0cm}

\usetikzlibrary{shapes.arrows}

\tikzset{%
	partial ellipse/.style args={#1:#2:#3}{%
		insert path={+ (#1:#3) arc (#1:#2:#3)}%
	}%
}%
\tikzstyle{style A}=[black, mark=diamond*, mark options={solid, fill=white, mark size=2.0pt}, solid]%
\tikzstyle{style B}=[color=blue, mark=*, mark options={solid, fill=white, mark size=1.5pt}, dotted]%
\tikzstyle{annotation}=[fill=white]%

\tikzstyle{diamond marker}=[mark=diamond*, mark options={solid, fill=white, mark size=2.0pt}]
\tikzstyle{triangle marker}=[mark=triangle*, mark options={solid, fill=white, mark size=2.0pt}]
\tikzstyle{square marker}=[mark=square*, mark options={solid, fill=white, mark size=1.3pt}]
\tikzstyle{circle marker}=[mark=*, mark options={solid, fill=white, mark size=1.5pt}]

\newcommand{\HW}[1]{{\color{red}{[HW: #1]}}}

\DeclareMathOperator*{\argmax}{arg\,max}

\title{Benchmarking End-to-end Learning of MIMO\\ Physical-Layer Communication}

\author{%
Jinxiang Song\IEEEauthorrefmark{1}, 
Christian H\"{a}ger\IEEEauthorrefmark{1},
Jochen Schr\"{o}der\IEEEauthorrefmark{2}, Tim O'Shea\IEEEauthorrefmark{3}, 
Henk Wymeersch\IEEEauthorrefmark{1}\\
\IEEEauthorrefmark{1}Department of Electrical Engineering, Chalmers University of Technology, Gothenburg, Sweden\\
\IEEEauthorrefmark{2}Department of Microtechnology and Nanoscience, Chalmers University of Technology, Gothenburg, Sweden\\
\IEEEauthorrefmark{3}Virginia Tech and DeepSig, Inc., Arlington, VA, USA
\thanks{This work was supported by the Knut and Alice Wallenberg Foundation, grant No.~2018.0090, and the Swedish Research Council under grant  No.~2018-0370. The work of C.~H\"ager was supported by the European Union's Horizon 2020 research and innovation programme under the Marie Sk\l{}odowska-Curie grant No.~749798. }
}

\begin{document}

\maketitle

\begin{abstract}
End-to-end data-driven machine learning (ML) of multiple-input multiple-output (MIMO) systems has been shown to have the potential of  exceeding the performance of engineered MIMO transceivers, without any a priori knowledge of communication-theoretic principles. In this work, we aim to understand to what extent and for which scenarios this claim holds true when comparing with fair benchmarks. We study closed-loop MIMO, open-loop MIMO, and multi-user MIMO and show that the gains of ML-based communication in the former two cases can be to a large extent ascribed to implicitly learned geometric shaping and bit and power allocation, not to learning new spatial encoders. For MU-MIMO, we demonstrate the feasibility of a novel method with centralized learning and decentralized executing, outperforming conventional zero-forcing. For each scenario, we provide explicit descriptions as well as open-source implementations of the selected neural-network architectures. 

\end{abstract}

\section{Introduction}
The ever-growing demand for higher data rates has led to the rapid development of wireless communication systems. One of the most important developments is multiple-input multiple-output (MIMO) transmission \cite{paulraj2004overview}, where information across multiple antenna elements is encoded using spatial-multiplexing or spatial-diversity schemes such that the throughput or reliability of  communication  can be improved in various channel conditions. Conventional MIMO communication systems are generally divided into closed-loop and open-loop. In open-loop systems, channel state information (CSI) is only available at the receiver, while in closed-loop systems, the transmitter also has access to CSI (either through explicit feedback or via channel reciprocity). 
Various algorithms have been proposed for both open-loop and closed-loop systems, including maximum-likelihood detection, zero-forcing (ZF) precoding, minimum mean-square-error (MMSE) equalization, space-time block coding, and singular value decomposition (SVD) with waterfilling \cite[Chapter 11]{WilWilBig16}.

Recent years have witnessed a resurgence of interest in machine-learning (ML) techniques for communication systems, where most works have focused on supervised ML for a \emph{specific functional block} such as modulation recognition \cite{o2016convolutional}, blind MIMO detection \cite{nguyen2018learning}, MIMO channel estimation \cite{he2018deep}, and channel decoding \cite{gruber2017deep}. These ML-based methods  have led  to  algorithms  that  either  perform  better  or  exhibit  lower complexity than model-based algorithms. 
In contrast to focusing on a specific functional block, \emph{end-to-end learning} has been proposed to optimize the transmitter and receiver jointly \cite{o2017introduction}. The workhorse of end-to-end learning is the autoencoder (AE), which employs two deep neural networks (NNs) to encode and decode messages into a learned latent representation which passes through a physical communication channel.
This method has been successfully applied to a wide variety of channels, including, e.g., linear wireless \cite{he2019model}, and nonlinear optical \cite{li2018achievable}. 
In cases where no channel model is available, a surrogate channel can first be learned \cite{Ye2018} or the transmitter can be designed as a reinforcement learning (RL) agent \cite{aoudia2018end}, which can operate even with limited reward feedback \cite{8891726}.

For MIMO communication, there has been limited treatment of AEs. To the best of our knowledge, the only directly related works are \cite{o2017physical, o2017deep}. 
In \cite{o2017physical}, open-loop and closed-loop MIMO was studied, leading to better performance than the selected benchmark methods. In the extension \cite{o2017deep}, finite quantization of the CSI was considered, which was demonstrated to improve performance in some conditions. 
While \cite{o2017physical, o2017deep} have shown promising performance of ML-based MIMO communication, the proposed AEs were trained under some nonstandard assumptions, specifically in terms of CSI availability at the receiver and power normalization at the transmitter, as explained below.

In this paper, we build on the approach proposed in \cite{o2017deep,o2017physical}, with the aim to 
better understand what performance can be achieved by ML-based MIMO solutions under realistic training assumptions, how and why they outperform standard benchmarks, and what limitations are imposed by NN architectures. Our main contributions in this work are as follows:
\begin{itemize}
    \item We revisit the MIMO systems in \cite{o2017deep,o2017physical} and evaluate the AEs under more standard training assumptions. In particular, while CSI in \cite{o2017deep,o2017physical} was assumed to be estimated at the receiver, it was not actually used as a receiver input. Also, power normalization was applied after the channel-matrix multiplication (cf.~\cite[Eqs.~(2), (3)]{o2017deep}), which cannot be done in practical systems. By contrast, our AEs always use the CSI as an additional receiver input and power normalization is performed prior to the channel. Moreover, reproducible open-source implementations of our AEs are also provided.

    \item We then explain some of the performance gains obtained by the trained AEs through the selection of improved baseline schemes compared to \cite{o2017deep,o2017physical}. In particular, for open-loop MIMO, we show that performance gains can be partially attributed to an implicit geometric shaping of the underlying signal constellation. For closed-loop MIMO, we use an SVD-based benchmark similar to \cite{o2017deep,o2017physical}, but augment it through additional bit and power allocation. This partially closes the performance gap to the AE, indicating that the ML-based solution learns to  implement similar functionalities in a data-driven fashion.

    \item Lastly, we consider a multi-user MIMO scenario, where a single multi-antenna transmitter sends information to multiple single-antenna users.\footnote{This scenario was suggested as a possible extension in \cite[Sec.~V]{o2017deep}.} For such a system, we extend the training methodology in \cite{o2017deep} to account for the joint loss function of all users and show that the resulting ML-based system achieves better performance compared to the considered baseline approach of applying transmitter ZF.

\end{itemize}

\subsection*{Notation}
We will use the following notations: $[a,b]^M$ is the $M$--fold Cartesian product of the $[a,b]$--interval. $\mathcal{CN}(\boldsymbol{x};\boldsymbol{\mu},\boldsymbol{\Sigma})$ denotes the distribution of a proper complex Gaussian random vector with mean $\boldsymbol{\mu}$, covariance matrix $\boldsymbol{\Sigma}$, evaluated in $\boldsymbol{x}$ ($\boldsymbol{x}$ may be omitted to represent the entire distribution). A matrix $\boldsymbol{X}$ is converted to a vector by stacking the columns, denoted by $\mathrm{vec}(\boldsymbol{X})$.

\section{Background and Baseline Schemes}

In this section, we describe the open-loop MIMO, closed-loop MIMO, and MU-MIMO systems under consideration and provide the benchmark transmitter and receiver algorithms. The channel at discrete time $k$ is denoted by $\boldsymbol{H}_k \in \mathbb{C}^{N_R \times N_T}$ for $N_R$ receive and $N_T$ transmit antennas. The channel is drawn from a stationary distribution $\boldsymbol{h}_k = \mathrm{vec}(\boldsymbol{H}_k) \sim p(\boldsymbol{h})$ and is assumed to be block fading with duration $N_B \geq N_T$. The transmitter can send sequences of messages belonging to a set $\mathcal{M}=\{ 1,2,\ldots, M\}$. The transmission rate is assumed to be fixed and forward error correcting coding is not considered. An average transmit power of $P_T$ is assumed.

\subsection{Open-loop MIMO}
\label{open-loop}
In open-loop systems, CSI is available at the receiver but not at the transmitter. Conventional transmit approaches include space-time block codes (STBCs), which are described next. 

The transmitter generates $L$ messages, maps each to a complex data symbol $s_{k,l}$ and then encodes $\boldsymbol{s}_k=[s_{k,1},\ldots,s_{k,L}]^{\mathsf{T}}$ using a STBC with rate $L/N_B \le 1$. The resulting $N_B$ coded vectors of length $N_T$ are $\boldsymbol{X}_k=[\boldsymbol{x}_{k,1},\ldots,\boldsymbol{x}_{k,N_B}]$, with the property that $\boldsymbol{X}^{\mathsf{H}}_k\boldsymbol{X}_k={P_T}\boldsymbol{I}_P$. If each of the $L$ complex data symbols corresponds to $\log_2(M)$ bits (i.e., one message), then the total bit rate is $r=L\log_2(M)/N_B$. In this paper, we restrict ourselves to the Alamouti code \cite{alamouti1998simple} (see also \cite[Fig.~6]{o2017deep}), where $N_B=2$, $L=2$, with $r=\log_2(M)$. 

The receiver observes
\begin{align}
    \boldsymbol{Y}_{k}=\boldsymbol{H}_k\boldsymbol{X}_{k} + \boldsymbol{N}_{k},
    \label{eq:OLMIMO-obs}
\end{align}
where $\mathrm{vec}(\boldsymbol{N}_{k}) \sim \mathcal{CN}(\boldsymbol{0},N_0\boldsymbol{I}_{N_R N_B})$ is i.i.d.~Gaussian noise. The receiver then applies maximum-likelihood detection to $\boldsymbol{Y}_{k}=[\boldsymbol{y}_{k,1},\ldots, \boldsymbol{y}_{k,P}]$, which can be achieved through low-complexity linear processing \cite{tarokh1999space}. Other (less complex) receiver approaches for open-loop MIMO include ZF and MMSE detection, which are not considered here.

\subsection{Closed-loop MIMO}
In closed-loop MIMO systems, the CSI is estimated at the receiver side, and then fed back to the transmitter. The most common approach is SVD-based MIMO, which we describe next.
    
    The block fading duration is irrelevant, but should be long enough to allow feedback and use of the CSI $\boldsymbol{H}_k$. The CSI is known to both transmitter and receiver, allowing both to compute the SVD
  $ \boldsymbol{H}_k = \boldsymbol{U}_k\boldsymbol{\Sigma}_k\boldsymbol{V}_k^\mathsf{H}$,  
where $\boldsymbol{\Sigma}_k=\text{diag}[\sigma_{k,1},\ldots, \sigma_{k,R_H}]$,  $\sigma_{k,1}\ge \sigma_{k,2} \ge \cdots \ge \sigma_{k,R_H}>0$, in which $R_H$ is the rank of $\boldsymbol{H}_k$. Correspondingly, $\boldsymbol{U}_k\in \mathbb{C}^{N_R\times R_H}$ and $\boldsymbol{V}_k\in \mathbb{C}^{N_T\times R_H}$ are truncated unitary matrices.

For each singular value $\sigma_{k,i}$, the transmitter chooses a constellation $\Omega_{i}$ from a set of available constellations as well as a transmit power $P_{T,i}\ge 0$. This selection can be based on the total message error rate according to
\begin{subequations}
\label{eq:svd_baseline}
\begin{align}
    \underset{\Omega_{i},P_{T,i}}{\text{maximize}}~~~ & \textstyle{\prod_{i=1}^{R_H}}(1-P_e(\Omega_{i},\gamma_i))\\
    \text{s.t.}~~~& \textstyle{\prod_{i=1}^{R_H}} |\Omega_{k,i}|=M,\\
     & \textstyle{\sum_{i=1}^{R_H}} P_{T,i}\le P_T,\\
     & \gamma_i = \frac{\sigma^2_{k,i}P_{T,i}}{N_0},
\end{align}
\end{subequations}
where $P_e(\Omega,\gamma)$ is the symbol error probability of constellation $\Omega$ under the specific receive SNR $\gamma$.  Hence, the rate is fixed to $r=\log_2(M)$. The corresponding symbol vector $\boldsymbol{s}_k = [s_{k,0}, s_{k,1},\cdots, s_{k,R_H}]^{\mathsf{T}}$ is precoded by $\boldsymbol{V}_k$, so that $\boldsymbol{x}_k=\boldsymbol{V}_k\boldsymbol{s}_k$, where $\mathbb{E}\{\Vert \boldsymbol{x}_k\Vert ^2\}\leq P_T$, is sent over the channel. 

The receiver observes $\boldsymbol{y}_k  = \boldsymbol{H}_k\boldsymbol{x}_k + \boldsymbol{n}_k$ and applies a combiner $\boldsymbol{U}^{\mathsf{H}}_k$, leading to the observation
\begin{align}
    \boldsymbol{\hat{y}}_k=\boldsymbol{U}^{\mathsf{H}}_k\boldsymbol{H}_k\boldsymbol{V}_k\boldsymbol{s}_k + \boldsymbol{U}^{\mathsf{H}}_k\boldsymbol{n}_k = \boldsymbol{\Sigma}_k\boldsymbol{s}_k+ \boldsymbol{U}^\mathsf{H}\boldsymbol{n}_k
\end{align}
where $ \boldsymbol{U}^\mathsf{H}\boldsymbol{n}_k$ has the same distribution as $\boldsymbol{n}_k$.  Maximum likelihood recovery of the transmitted messages is straightforward, since $\boldsymbol{\Sigma}_k$ is a diagonal matrix.

\subsection{MU-MIMO}
\label{sec:mu-mimo}
We consider a downlink MU-MIMO system where there are a transmitter with $N_T$ antennas and $N_R$ receivers each with one antenna, where $N_T\geq N_R$. To eliminate the interference among different users, various algorithms including linear and non-linear precoding have been proposed. In this paper, we consider a linear precoding scheme referred to as transmitter ZF, which we describe next.
 
    Similar to the closed-loop MIMO case, the block fading duration is irrelevant, but should be long enough to allow feedback and use of the CSI. The local CSI $\boldsymbol{h}^{\mathsf{T}}_{k,i}\in \mathbb{C}^{1\times N_T}, i=1,\ldots,N_R$ is estimated at each user and fed back to the transmitter. Thus, the transmitter has knowledge of  the full CSI
    $\boldsymbol{H}_k=[\boldsymbol{h}_{k,1},\ldots,\boldsymbol{h}_{k,N_R} ]^\mathsf{T}$, 
while each receiver only has access to the local CSI.

    The transmitter encodes $N_R$ messages $\boldsymbol{s}_k = [s_{k,1},\cdots,s_{k,N_R}]^\mathsf{T}$ with the pseudo-inverse of the channel matrix $\boldsymbol{H}_k^\dagger$, so that $ \boldsymbol{x}_k=\alpha \boldsymbol{H}_k^\dagger\boldsymbol{s}_k$ is sent over the channel, where $\boldsymbol{H}_k^\dagger =\boldsymbol{H}_k^{\mathsf{H}}(\boldsymbol{H}_k\boldsymbol{H}_k^{\mathsf{H}})^{-1}$ and $\alpha$ is set to ensure that $\mathbb{E}\{\Vert \boldsymbol{x}_k\Vert ^2\}\leq P_T$. If each of the messages corresponds to $\log_2(M)$ bits, the sum-rate of the system is $r = N_R\log_2(M)$.
    
    Each user $i$ observes 
    $y_{k,i} = \boldsymbol{h}^{\mathsf{T}}_{k,i} \boldsymbol{x}_k + n_{k,i}=\alpha s_{k,i} + n_{k,i},$
from which $s_{k,i}$ can be recovered with low-complexity maximum-likelihood detection.

\section{Autoencoders for MIMO Systems}

In this section, we first describe the idea behind AE-based communication for a single-input single-output system. We then describe the AE implementation for the open-loop, closed-loop, and MU-MIMO systems. For all scenarios, the transmitters are denoted by $f_\tau(\cdot)$ and the receivers by $f_\rho(\cdot)$. 

\subsection{AE-based Communication Systems}
\label{sec:siso_ae}

AE-based end-to-end learning was proposed in \cite{o2017introduction}. 
For a single-input single-output system, the transceiver is implemented by a pair of multi-layer NNs $f_\tau: \mathcal{M}\to \mathbb{C}$ and $f_\rho:\mathbb{C}\to[0,1]^M$,  where $\tau$ and $\rho$ are the transmitter and receiver parameters.

\subsubsection{Transmitter}
Given a message $m_k\in\mathcal{M}$, the transmitter generates 
$x_k=f_\tau(m_k)$, where an average power constraint $\mathbb{E}\{|x_k|^2\}\leq P_T$ is enforced by a normalization layer. The message $m_k$ is assumed to be encoded to an $M$--dimensional ``one-hot'' vector $\boldsymbol{l}_k \in \{0,1 \}^M$, where the $m$--th element is $1$ and all the others are $0$.

\subsubsection{Receiver}
The complex symbol $x_k$ is sent over the channel, and the receiver processes the received symbol $y_k$ by first generating an $M$-dimensional probability vector $\boldsymbol{q}_k=f_\rho(y_k)$, where the components of $\boldsymbol{q}_k$ can be interpreted as the estimated posterior probabilities of the messages. Finally, the transmitted message is estimated according to $\hat{m}_k=\argmax_m[\boldsymbol{q}_k]_m$, where $[\boldsymbol{x}]_m$ returns the $m$-th element of $\boldsymbol{x}$.

\subsubsection{End-to-end learning}
To optimize the transmitter and receiver parameters, it is important to have a suitable optimization criterion. Due to the fact that optimization relies on the empirical computation of gradients, a criterion like symbol error rate (SER) $\mathrm{Pr}\{\hat{m}_k\neq m_k\}$ cannot be used directly.
Instead, a commonly used criterion is the categorical cross-entropy loss function ($\mathcal{J}_{\text{CE}}$) defined by
\begin{equation}
\label{eq:ce_loss}
    \mathcal{J}_{\text{CE}}(\tau, \rho)=-\mathbb{E}_{m_k,y_k}\{ \log [f_{\rho}(y_k)]_{m_k}\},
\end{equation}
where the dependence of $J_{\text{CE}} (\tau,\rho)$ on $\tau$ is implicit through the distribution of the channel output $y_k$, which is a function of the channel input $f_{\tau}(m_k)$. In practice, $\mathcal{J}_\text{CE}$ is usually approximated via Monte-Carlo estimation according to $\hat{\mathcal{J}}_\text{CE} = \frac{1}{B_S}\sum_{i=1}^{B_S}\{ \log [f_{\rho}(y_k)]_{m_k}\}$, where $B_s$ is the batch size. And optimization of the NNs can be performed by minimizing $\hat{\mathcal{J}}_{\text{CE}}$ through the widely used  Adam optimizer \cite{kingma2014adam}.

\subsection{Open-loop MIMO AE}

\label{open-loop-mimo}
\begin{figure}[t]
    \centering
    \includegraphics[width=\columnwidth]{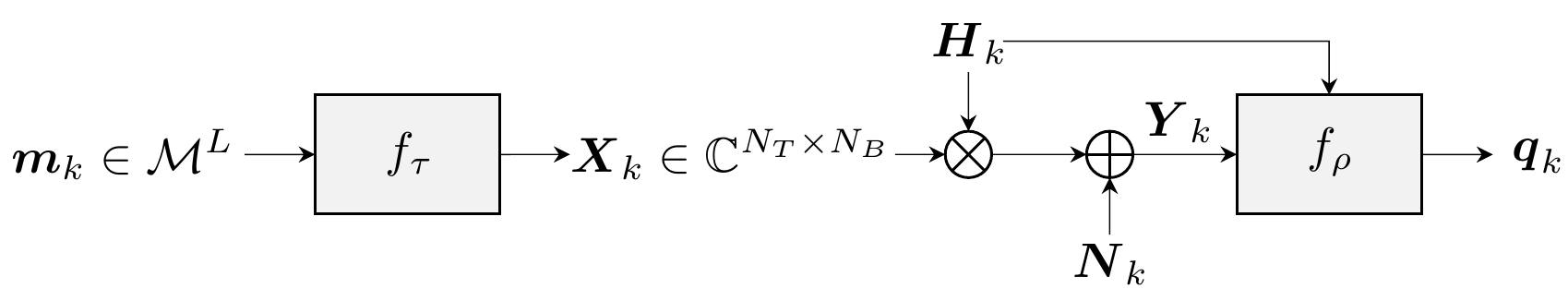}
    \caption{Open-loop MIMO channel AE, where the transmitter learns a rate $L/N_B$ code without CSI, while the receiver learns a decoder in the presence of CSI. The channel is drawn i.i.d.~from the underlying distribution.}
    \label{fig:open-loopSYSTEM}
\end{figure}

For an open-loop MIMO system with CSI available to the receiver, the AE implementation is visualized in Fig.~\ref{fig:open-loopSYSTEM}. 
The transmitter $f_\tau: \mathcal{M}^L\to \mathbb{C}^{N_T\times N_B}$ maps $L$ consecutive messages $\boldsymbol{m}_k = (m_1, \ldots, m_L) \in \mathcal{M}^L$ into $N_B$ coded vectors according to
\begin{equation}
   \boldsymbol{X}_k = [\boldsymbol{x}_{k,1}, \ldots, \boldsymbol{x}_{k,
   N_B}]= f_\tau(\boldsymbol{m}_k),
\end{equation}
where $ \boldsymbol{x}_{k,p}$, $p=1, \ldots,N_B$, is a column vector of length $N_T$. An average power constraint according to $\sum_{p=1}^{N_B} \mathbb{E}\{\Vert \boldsymbol{x}_{k,p} \Vert ^2\} \leq N_B P_T$ is enforced. Inside $f_\tau(\cdot)$, an encoding of $\boldsymbol{m}_k$ to an $M^L$-dimensional one-hot vector is used.

The receiver $f_{\rho}: \mathbb{C}^{N_R\times N_B} \times \mathbb{C}^{N_R\times N_T} \to [0,1]^{M^L}$ observes $\boldsymbol{Y}_{k}=[\boldsymbol{y}_{k,1},\cdots \boldsymbol{y}_{k,N_B}]$ as in \eqref{eq:OLMIMO-obs} and generates a probability vector $\boldsymbol{q}_k\in[0,1]^{M^L}$ according to 
\begin{align}
\label{eq:NN_decoder}
    \boldsymbol{q}_k=f_\rho(\boldsymbol{Y}_{k}, \boldsymbol{H}_k),
\end{align}
in which the CSI $\boldsymbol{H}_k$ is concatenated to the observations $\boldsymbol{Y}_k$ and then provided to the receiver.\footnote{The CSI is first converted  into a real-valued vector of length $2 N_R N_T$, and then concatenated to the observations, which are also converted into a real-valued vector. } Finally, the transmitted message  is estimated as $\hat{m}_k=\argmax_m[\boldsymbol{q}_k]_m$.

Note that while the transmitter does not have access to \emph{instantaneous} CSI in the learning process, it can obtain knowledge of the CSI distribution $p(\boldsymbol{h})$, i.e., \emph{statistical} CSI.

\subsection{Closed-loop MIMO AE}
\begin{figure}[t]
     \centering
     \includegraphics[width=1\columnwidth]{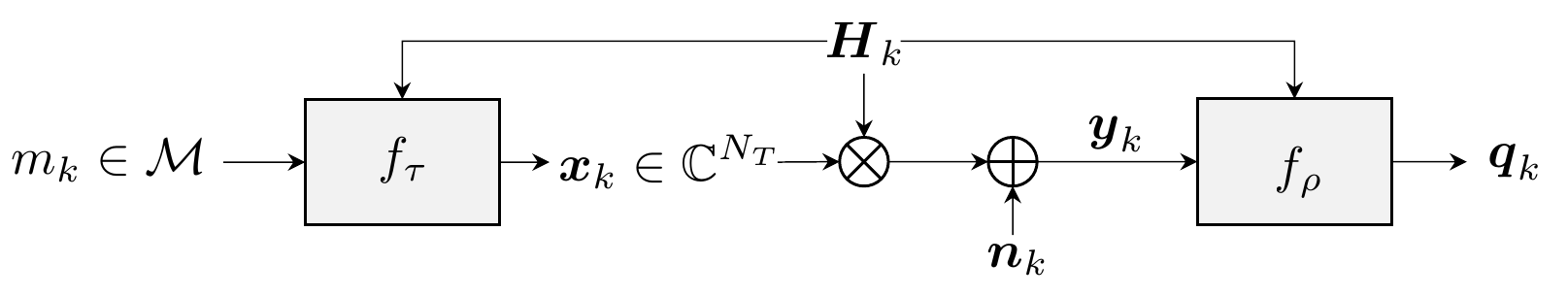}
     \vspace{-0.1cm}
     \caption{Closed-loop MIMO AE, in which both transmitter and receiver have access to CSI.  }
     \label{fig:closed-loopSYSTEM}
 \end{figure}

In the closed-loop MIMO system, CSI is known to both transmitter and receiver.  The AE is implemented as shown in Fig.~\ref{fig:closed-loopSYSTEM}. To provide the transmitter with CSI, the transmitter is of the form $f_{\tau}:\mathcal{M}\times \mathbb{C}^{N_R\times N_T} \to \mathbb{C}^{N_T\times1} $, yielding complex vectors
    $\boldsymbol{x}_k=f_\tau(m_k, \boldsymbol{H}_k)$
in which the transmitter is provided with the message $m_k$ as well as the CSI $\boldsymbol{H}_k$. A one-hot encoding is used with vectors of length $M$ and a normalization layer ensures  $\mathbb{E}\{\Vert \boldsymbol{x}_{k} \Vert ^2\} \leq P_T$

The receiver $f_{\rho}: \mathbb{C}^{N_R\times 1} \times \mathbb{C}^{N_R\times N_T} \to [0,1]^{M}$ observes $\boldsymbol{y}_k = \boldsymbol{H}_k \boldsymbol{x}_k + \boldsymbol{n}_k$, and similar to the open-loop MIMO case, the transmitted message  is estimated as $\hat{m}_k=\argmax_m[\boldsymbol{q}_k]_m$, where $\boldsymbol{q}_k=f_\rho(\boldsymbol{y}_{k}, \boldsymbol{H}_k)$ is a probability vector obtained in the same way as in \eqref{eq:NN_decoder}.

\subsection{MU-MIMO AE}
\begin{figure}[t]
    \centering
    \includegraphics[width=\columnwidth]{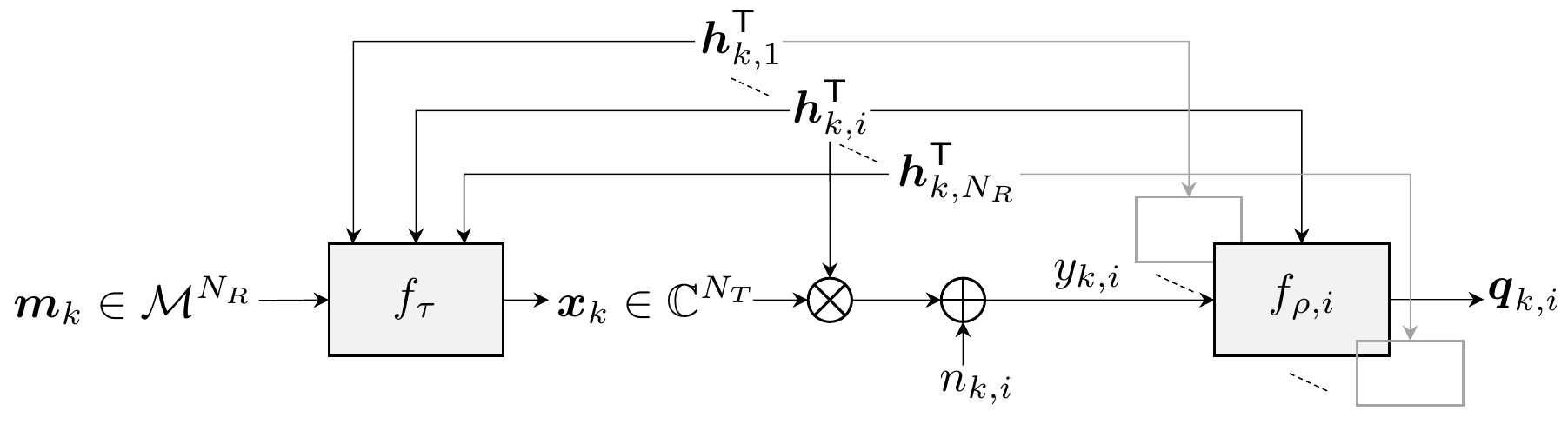}
    \vspace{-0.1cm}
    \caption{MU-MIMO AE, in which the transmitter encodes messages for the individual users, based on full CSI, while each user observes only a local measurement and local CSI. }
    \label{fig:mu-mimoSYSTEM}
\end{figure}

For a MU-MIMO system, each receiver only has  access to the local CSI, while the transmitter has knowledge of the full CSI. The AE implementation is visualized in Fig.~\ref{fig:mu-mimoSYSTEM}. 

The transmitter $f_\tau: \mathcal{M}^{N_R} \times  \mathbb{C}^{N_R\times N_T} \to \mathbb{C}^{N_T\times 1}$ maps individual messages $m_{k,i}\in \mathcal{M}$ for each user $i=1,\cdots, N_R$ to $N_T$ complex symbols. With $\boldsymbol{m}_k=[m_{k,1},\cdots,m_{k, N_R}]^{\mathsf{T}}$,  
\begin{align}
    \boldsymbol{x}_k = f_\tau(\boldsymbol{m}_k, \boldsymbol{H}_k)
\end{align}
where an average power constraint $\mathbb{E}\{\Vert \boldsymbol{x}_k\Vert ^2\}\leq P_T$ is enforced. One-hot encoding of $\boldsymbol{m}_k$ to a vector of length $M^{N_R}$ is applied.

The $N_R$ receivers are implemented by $N_R$ NNs, each with different parameters, of the form $f_{\tau,i}:\mathbb{C} \times \mathbb{C}^{N_T} \to [0,1]^M$. 
Each user $i$ observes $y_{k,i} = \boldsymbol{h}^\mathsf{T}_{k,i}\boldsymbol{s}_k + n_{k,i}$ and generates a probability vector $\boldsymbol{q}_{k,i}\in[0,1]^M$ according to 
\begin{align}
    \boldsymbol{q}_{k,i} = f_{\rho, i}(y_{k,i}, \boldsymbol{h}_{k,i}),
\end{align}
in which the receiver is provided with its observation $y_{k,i}$ as well as the local CSI $\boldsymbol{h}_{k,i}$. Then, the transmitted message for the $i$--th user is estimated as $m_{k,i}=\argmax_m[\boldsymbol{q}_{k,i}]_m$.

In order to train the MU-MIMO AE, the cross-entropy loss function defined in \eqref{eq:ce_loss} cannot be used directly, as we now have several receivers that need to be optimized. Instead, we apply a joint loss function 
\begin{align}
\label{eq: joint_ce}
& \mathcal{J}_{\text{CE}}(\tau, \rho_1, \cdots, \rho_{N_R}) =\\
& -\sum_{i=1}^{N_R}\mathbb{E}_{m_{k,i},\boldsymbol{y}_{k,i},\boldsymbol{h}_{k,i}}\left\{  \log[f_{\rho_i}(\boldsymbol{y}_{k,i})]_{m_{k,i}} \right\}, \nonumber
\end{align}
which can again be optimized using the Adam optimizer.

\section{Numerical Results}

In this section, we provide simulation results for the open-loop, closed-loop, and MU-MIMO systems.\footnote{The complete source code to reproduce all results in this paper is available at \url{https://github.com/JSChalmers/DeepLearning_MIMO.git}}  For all 3 cases, the channel model is assumed to be Rayleigh fading, i.e., $p(\boldsymbol{h})=\mathcal{CN}(\boldsymbol{h};\boldsymbol{0},\boldsymbol{I}_{N_R N_T})$, and the system performance is measured in terms of SER as a function of the average signal-to-noise ratio (SNR), defined as $\mathrm{SNR}=P_T/N_0$.


\subsection{NN Architectures and Training Procedure}

All AEs are implemented as multi-layer fully-connected NNs, where the ReLU function is chosen as the activation function. Detailed NN parameters are summarized in Table ~\ref{tab:network_parameters}, where the shown parameters for the open-loop case are only used for $M=4$. For $M=16$, the number of hidden layers for both the transmitter and receiver NNs is increased to 5. All AEs are trained by using the Adam optimizer \cite{kingma2014adam} with a learning rate $0.001$ and batch size $2048$.

\subsection{Open-loop MIMO}

We use the parameters $N_T=2, N_R=1, N_B=2, L=2$, and $M\in \{ 4,16\}$. The training is performed at $\mathrm{SNR}=15~\text{dB}$ for $M=4$ and $\mathrm{SNR}=18~\text{dB}$ for $M=16$ over $3\,200\,000$ random channel realizations.

\begin{table}[t]
\setlength{\tabcolsep}{0.6em}
\scriptsize
\centering
\vspace{0.15cm}
\caption{Neural-network parameters for (i) closed-loop, (ii) open-loop, and (iii) MU-MIMO systems}
\begin{tabular}{c|c|ccc|ccc}
\toprule
& & \multicolumn{3}{c}{transmitter $f_\tau$}   & \multicolumn{3}{|c}{receiver $f_\rho$} \\
 \midrule
& layer    & input  & hidden& output    & input & hidden     & output  \\ 
\midrule
\multirow{3}{*}{(i)}&\#~of layers    & -  & $3$ & -    & - & $3$     & -  \\ 
&\#~of neurons   & $M$ & 64   & $8$      & $12$     & $512$      & $M$          \\ 
&act.~function & - & ReLU & linear & -     & ReLU    & softmax    \\ 
\midrule
\multirow{3}{*}{(ii)}&\#~of layers    & - & $2$  & -    & - & $3$     & -  \\ 
&\#~of neurons   & $M+8$ & $256$   & $8$      & $16$     & $2048$      & $M$          \\ 
&act.~function & - & ReLU & linear & -     & ReLU    & softmax    \\ 
\midrule
\multirow{3}{*}{(iii)}&\#~of layers    & - & $3$  & -    & - & $3$     & -  \\ 
&\#~of neurons   & $M+8$ & $512$   & $8$      & $6$     & $256$      & $M$          \\ 
&act.~function & - & ReLU & linear & -     & ReLU    & softmax    \\ 
\bottomrule
\end{tabular}
\label{tab:network_parameters}
\end{table}

\begin{figure}[t]
    \centering
    \footnotesize%
\renewcommand{\figwidth}{6.5cm}
\renewcommand{\figheight}{4.0cm}
\begin{tikzpicture}%
	\tikzstyle{annotation}=[fill=white]%
	\begin{semilogyaxis}[%
	    scale only axis,
		xlabel=SNR (dB), ylabel=symbol error rate,
		xmin=0, xmax=22.5,%
		ymin=1e-4, ymax=1,%
		grid=both,%
 		width=\figwidth, height=\figheight,%
		legend cell align={left},%
		label style={font=\scriptsize},
		legend style={at={(0.02,0.2)},anchor=west,nodes={scale=0.9, transform shape}}%
	]%
	
	\addplot +[thick, color=blue, solid, mark=square*,mark options={scale=1,solid}] coordinates {(0, 3.57e-1) (2,2.413e-01)(4,1.455e-01) (6,8.24e-02) (8,4.170e-02)
    (10,1.9756e-02) (12,8.866e-03) (14,3.816e-03) (16, 1.602e-03)
    (18,6.610e-04)(20, 2.63000e-04 ) (22,9.80e-05) };
    \addlegendentry{Baseline} 
	    \addplot +[thick, color=green!50!black, solid, mark=circle*,mark options={scale=1,solid}] coordinates {(0,0.8551) (2, 0.7749) 
	(4, 0.66) (6, 0.5244)  (8, 0.3720)
    (10, 0.2389)  (12, 0.1385) (14, 0.0704)
    (16, 0.0344) (18, 0.0164)
    (20,  0.007) (22, 0.0031) 
    };
    \addlegendentry{Baseline w/ geom.~shaping} 
	  	\addplot +[thick, color=red, dashed, mark=triangle*,mark options={scale=1,solid}] coordinates {(0, 0.84780599) (2,0.763)(4,0.6454388) (6,0.50415365)
    (8,0.3567487 ) (10, 0.22659245) (12,0.12942448) (14, 0.0680507)
    (16, 0.03287109)(18, 0.01507943) (20,0.00656771 ) (22, 0.00233984)};
    \addlegendentry{Autoencoder}

		\addplot +[thick, color=red, dashed, mark=triangle*,mark options={scale=1,solid}] coordinates {(0,0.353) (2, 2.3849e-1) 
	(4, 1.4542e-1) (6, 8.02e-2)  (8, 4.0724e-2)
    (10, 1.9133e-2 )  (12, 8.582e-3) (14, 3.724e-3)
    (16, 1.5412e-3) (18, 6.596e-4)
    (20,  2.802e-4) (22, 1.093669e-4) 
    };
	
	\addplot +[thick, color=blue, solid, mark=square*,mark options={scale=1,solid}] coordinates {(0, 0.7048) (2,0.6478)(4,0.5866) (6,0.5291)
    (8, 0.4721) (10, 0.4025) (12,0.329) (14, 0.2525)
    (16, 0.1784)(18, 0.114) (20,0.06441) (22, 0.0353)};

	\draw[black] (11.5,9e-3) [partial ellipse=-90+30+360:90-30:.07cm and 0.3cm];%
	\node[fill=white, inner sep=0.4pt] (rrd) at (12.5, 3.0e-2) {\scriptsize\color{black}$M=4$};%
	
	\draw[black] (15,1.05e-1) [partial ellipse=90+30:-90-30:.08cm and 0.65cm];%
	\node[fill=white, inner sep=0.4pt] (norrd) at (17.5, 0.4) {\scriptsize\color{black} $M=16$};%

	\end{semilogyaxis}
	\end{tikzpicture}%
    \vspace{-0.1cm}
    \caption{SER of the open-loop MIMO AE and the baseline scheme consisting of standard $M$-QAM signal constellations, an Alamouti STBC, and a maximum-likelihood receiver. The improved baseline for $M=16$ uses a geometrically-shaped signal constellation.}
    \label{fig:alamouti}
\end{figure}
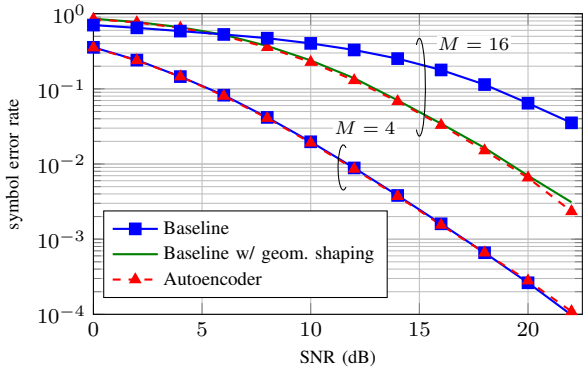

Fig.~\ref{fig:alamouti} shows the achieved SER results over a range of SNRs. As a reference, the performance of the baseline scheme described in Sec.~\ref{open-loop} is also shown. For $M=4$,  the AE achieves very similar performance to the baseline scheme, indicating that the combination of a QPSK constellation and Alamouti STBC is near-optimal in this case. For  $M=16$, the AE outperforms the baseline scheme significantly at high SNR when standard $16$-QAM is used as the signal constellation. In order to improve the baseline, we also used a geometrically-shaped signal constellation for $M=16$, which was obtained by training a standard single-input single-output AE (see Sec.~\ref{sec:siso_ae}) over an AWGN channel at $\mathrm{SNR}=12~\text{dB}$. When this geometrically-shaped constellation is used instead, the baseline scheme has essentially the same performance as the AE-based approach, indicating that the AE learns to perform a joint optimization over the signal constellation and STBC.

\begin{figure}[t]
    \centering
    \includegraphics[width=0.9\columnwidth]{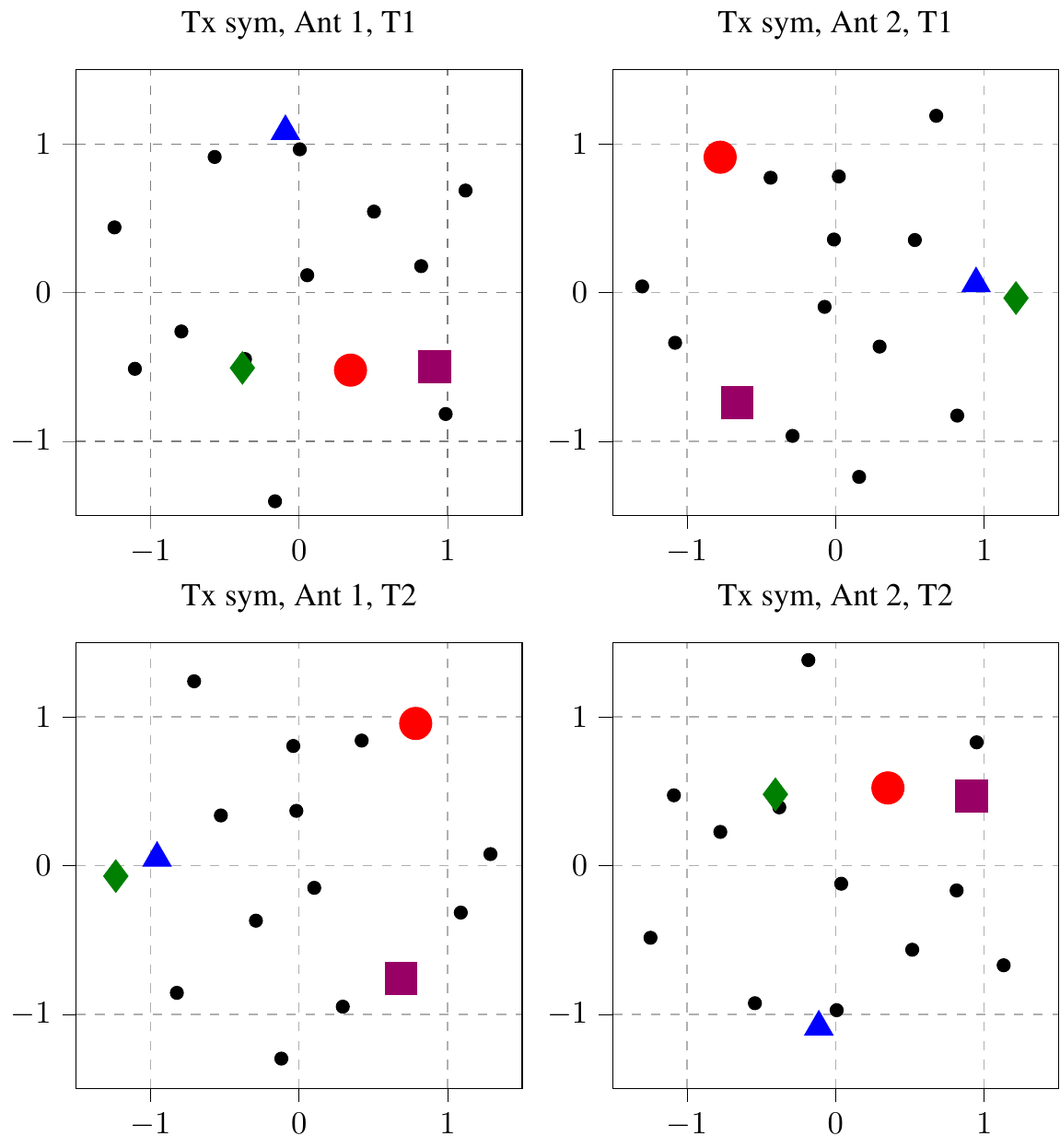}
    \caption{Learned transmitted symbols of the open-loop MIMO AE for $M=4$. (a) first antenna at time $p=1$, (b) second antenna at $p=1$, (c) first antenna at $p=2$ and (d) second antenna at $p=2$. Constellation points for $4$ out of $16$ messages are highlighted with colored markers.} 
    \label{fig:learnt_const}
\end{figure}

Fig.~\ref{fig:learnt_const} visualizes the learned transmitted symbols for $M=4$. Particularly, the constellation points for $4$ out of $M^L=16$ individual messages  are highlighted by different markers. From these plots, one can observe that the learned constellation follows a very similar pattern as the Alamouti scheme, in the sense that the symbols in subplot 1 are symmetric with respect to subplot 4 along the x-axis, while subplot 2 is symmetric with respect to subplot 3 along the y-axis.

\subsection{Closed-loop MIMO}

We use the parameters $N_T=2, N_R=2$, and $M=16$, corresponding to rate $r=4$. Training is performed at a fixed $\mathrm{SNR}=15~\text{dB}$ over $40\,960\,000$ random channel realizations. Different from the open-loop case, we notice that a lot more data samples are required for converging to a good solution.

\begin{figure}[t]
    \centering
    \footnotesize%
\renewcommand{\figwidth}{6.5cm}
\renewcommand{\figheight}{4.0cm}
\begin{tikzpicture}%
	\tikzstyle{annotation}=[fill=white]%
	\begin{semilogyaxis}[%
	    scale only axis,
		xlabel=SNR (dB), ylabel=symbol error rate,
		xmin=0, xmax=20.5,%
		ymin=4.5e-6, ymax=1,%
		grid=both,%
 		width=\figwidth, height=\figheight,%
		label style={font=\scriptsize},
		legend cell align={left},%
		legend style={at={(0.02,0.25)},anchor=west,nodes={scale=0.9, transform shape}}%
	]%

	\addplot +[thick, color=blue, solid, mark=square*, mark options={scale=1,solid }] coordinates {(0,0.51663) (2,0.42939) 
	(4, 0.34221) (6,0.26423)  (8, 0.1949)
   (10, 0.14124 )  (12,0.09486)
    (14, 0.06518)  (16, 0.04256) 
    (18, 0.02639)(20,  0.0175)  (22, 0.0112) (24, 0.0062)(26, 0.0046)
    };
	\addlegendentry{SVD w/ QPSK} 
	
	\addplot +[thick, color=green!50!black, solid, mark=circle*,mark options={scale=1,solid}] coordinates {(0, 4.037860e-01) (2,3.083340e-01)(4,2.132100e-01) (6,1.382575e-01) (8,7.183750e-02)
    (10,3.136500e-02) (12,1.133280e-02) (14,3.339900e-03) (16, 7.619000e-04)
    (18,1.655000e-04)(20, 2.330000e-05 ) (22,3.010000e-06) (24,3.700000e-07) (26,6.000000e-08)};
	\addlegendentry{SVD w/ bit and power allocation} 
	
	\addplot +[thick, color=red, dashed, mark=triangle*,mark options={scale=1,solid}] coordinates {(0, 0.51663) (2,4.133340e-01)(4,2.732100e-01) (6,1.682575e-01) (8,7.383750e-02)
    (10,2.336500e-02) (12,7.133280e-03) (14,1.339900e-03) (16, 3.619000e-04)
    (18,5.655000e-05)(20, 6.330000e-06 ) (22,1.010000e-06) (24,1.700000e-07) (26,2.000000e-08)};
	\addlegendentry{Autoencoder} 

	\end{semilogyaxis}
	\end{tikzpicture}%
    \vspace{-0.1cm}
    \caption{SER of the closed-loop MIMO AE for $M=16$ and the baseline scheme consisting of a QPSK constellation, SVD-based signal processing, and a maximum-likelihood receiver. The improved baseline uses bit and power allocation assuming BPSK, QPSK, 8-QAM, and 16-QAM constellations. }
    \label{fig:svd-ser}
\end{figure}
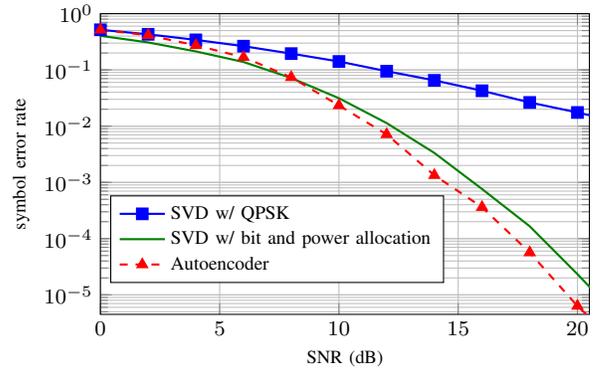

As a baseline, we simulate the performance of the SVD-based approach, in which the $2\times2$ MIMO channel is parallelized into two sub-channels. We first consider the same baseline as in \cite{o2017deep}, where equal power is used at each antenna, and both streams use QPSK modulation. However, depending on the channel realization, the two individual sub-channels will have different link quality, and bit and power allocation is usually used to improve the overall system performance. To that end, an improved baseline scheme was simulated by solving \eqref{eq:svd_baseline} using exhaustive search assuming that the set of available signal constellations is BPSK, QPSK, $8$--QAM, and $16$--QAM. Fig.~\ref{fig:svd-ser} shows the achieved SER for both the AE implementation as well as the two SVD-based approaches. While the AE achieves better performance than the SVD-based approach including bit and power allocation, we believe that the baseline could be further improved by changing the available set of signal constellations. These results indicate that for a fixed targeted transmission rate, the closed-loop MIMO AE learns to implicitly perform a combination of constellation shaping, bit allocation, and power allocation jointly.

\subsection{MU-MIMO}

We use the parameters $N_T=2, N_R=2$, and $M=16$, corresponding to a sum rate $r=4$. Training is performed at a fixed $\mathrm{SNR}=15~\text{dB}$ over $6\,400\,000$ channel realization. Compared to the previous two cases, there are now three different NNs, one corresponding to the transmitter and two to the individual users, where the same network architecture is used for both users.

\begin{figure}[t]
    \centering
    \footnotesize%
\renewcommand{\figwidth}{6.5cm}
\renewcommand{\figheight}{4.0cm}
\begin{tikzpicture}%
	\tikzstyle{annotation}=[fill=white]%
	\begin{semilogyaxis}[%
	    scale only axis,
		xlabel=SNR (dB), ylabel=symbol error rate,
		xmin=0, xmax=22.5,%
		ymin=5e-4, ymax=1,%
		grid=both,%
 		width=\figwidth, height=\figheight,%
		legend cell align={left},%
		label style={font=\scriptsize},
		legend style={at={(0.02,0.2)},anchor=west,nodes={scale=0.9, transform shape}}%
	]%

\addplot [thick, color=blue, solid, mark=square*,mark options={scale=1,solid}]
table {%
0 0.46261621596
2 0.39535978728
4 0.32294428003
6 0.25086304974
8 0.18474609132
10 0.13045413061
12 0.08914125202
14 0.0589115395
16 0.0389452427
18 0.02513767375
20 0.0159980387
22 0.0103117189
24 0.0064150058
26 0.0041246665
};
\addlegendentry{Zero-forcing precoder}
	\addplot [thick, color=red, dashed, mark=triangle*,mark options={scale=1,solid}]
table {%
0 0.410484375
1 0.37421875
2 0.33553125
3 0.2950703125
4 0.2585390625
5 0.220125
6 0.1790078125
7 0.1389921875
8 0.10725
9 0.0786718749999999
10 0.05459375
11 0.0351015625
12 0.0219140625000001
13 0.0132265625
14 0.00763281250000003
15 0.00410156250000004
16 0.00307031250000001
17 0.0018359375
18 0.00125781250000001
19 0.0009765625
20 0.000906249999999997
21 0.000773437500000029
22 0.000765624999999992
23 0.000765624999999992
};
\addlegendentry{Autoencoder}

	\end{semilogyaxis}
	\end{tikzpicture}%
    \vspace{-0.1cm}
    \caption{SER of the MU-MIMO AE for $M=16$, $2$ transmit antennas, and $2$ users and the baseline scheme consisting of a QPSK constellation, ZF transmitter, and a maximum-likelihood receiver. }
    \label{fig:mu-mimo}
\end{figure}
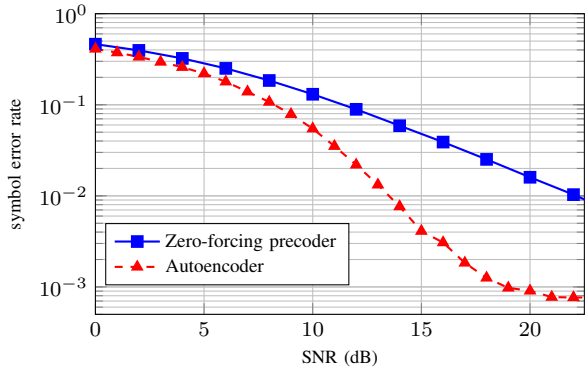

Fig.~\ref{fig:mu-mimo} shows the achieved SER results for the MU-MIMO AE as well the performance of baseline approach with transmitter ZF described in Sec.~\ref{sec:mu-mimo}. It can be seen that the AE-based MU-MIMO scheme achieves significantly better performance than the ZF-based approach, indicating that for independently operating receivers, the AE can learn  novel encodings. Further improving the baseline scheme for the MU-MIMO case is part of ongoing research. 

\subsection{Training Complexity}

\begin{figure}[t]
    \centering
    \includegraphics[width=0.9\columnwidth]{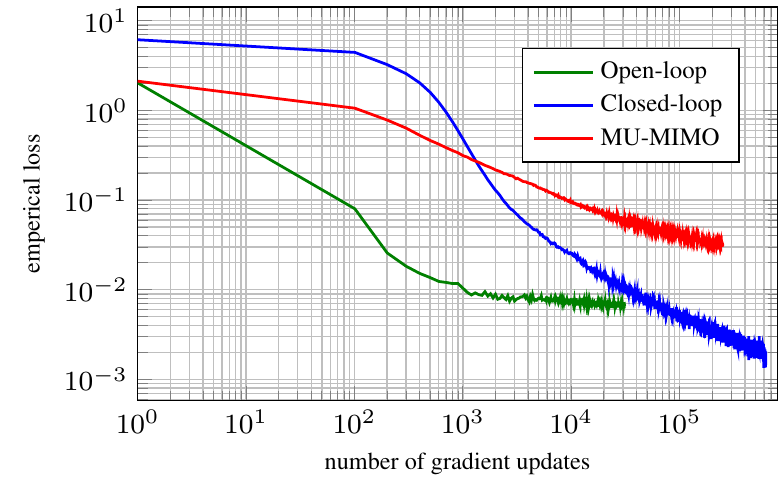}
    \vspace{-0.1cm}
    \caption{Evolution of $\mathcal{J}_{\text{CE}}(\cdot)$ versus number of gradient updates. }
    \label{loss_evolution}
\end{figure}
Fig.~\ref{loss_evolution} visualizes the evolution of training loss as a function of the number of gradient updates. Surprisingly, for the MU-MIMO case, even though the training process does not appear to have fully converged yet, the AE still outperforms the standard transmitter ZF approach quite significantly. We believe that more training data or gradient updates can lead to even better performance.

\section{Conclusion}

In this work, we have evaluated several AE-based MIMO communication systems, in order to quantify gains with respect to fair benchmarks. The systems under consideration were  open-loop MIMO, closed-loop MIMO and MU-MIMO, for which the AE provides optimized mappings from messages to transmit vectors, as well as optimized detectors. For open-loop and closed-loop MIMO, we have shown that the gains of the AE compared to the baselines can be partially attributed to geometric constellation shaping and optimized bit and power allocation. For MU-MIMO, we proposed a novel decentralized AE structure that is demonstrated to outperform ZF precoding. For each of these systems, we have provided open-source implementations.

Nevertheless, there are several important limitations of MIMO AEs, which deserve further study:
\begin{itemize}
    \item \emph{Training complexity:} The AE requires a very large amount of training data, with large batch sizes, in order to converge to a good solution. Smart selection of specific channel realizations can improve convergence speed. 
    \item \emph{Scalability:}  With more transmit and receive antennas or more users, the complexity scaling of the NN (e.g., in terms of layers) is currently unknown and the employed one-hot encoding scheme causes input and output sizes to grow exponentially with number of antennas and rate. Alternative embeddings \cite{rodriguez2018beyond} or multi-hot sparse categorical cross entropy could help alleviate the latter issue.  Both these issues affect training convergence (due to more trainable parameters) and runtime computational complexity. 
    \item \emph{Rate adaptation:} The considered AEs have a fixed data rate, which limits possibilities for rate adaptation. New NN architectures are needed to provide rate-adaptive transmission. 
\end{itemize}

\balance
\bibliographystyle{IEEEtran}
\bibliography{references} 
\label{references}
\end{document}